\documentclass[onecolumn,showpacs,showkeys,preprintnumbers,amsmath,amssymb,nofootinbib]{revtex4-1}
\usepackage[latin1]{inputenc}       
\usepackage[english]{babel}
\usepackage{bm}                        
\usepackage{graphicx,color}
\usepackage{amsmath}
\usepackage{epsfig}
\usepackage{gensymb}                
\usepackage{dcolumn}                

\begin{document}
\def\be{\begin{equation}}
\def\ee{\end{equation}}
\def\lb{\label}

\draft 

\preprint{SSC/2017}

\title[JOURNAL OF APPLIED PHYSICS]{Influence of the Terbium film size for the magnetocaloric effect at low temperatures}


\author{V.D. Mello$^{1}$\footnote{Phone: +55 84 3315 2196}}
\email{vambertodias@uern.br}
\author{D.H.A.L. Anselmo$^{2}$}
\author{M.S. Vasconcelos$^{3}$}
\author{N.S. Almeida$^{1}$}

\affiliation{$^{1}$Departamento de F\'{\i}sica, Universidade do Estado do Rio Grande do Norte, Mossor\'o - RN 59625-620, Brazil}
\affiliation{$^{2}$Departamento de F\'{\i}sica Teórica e Experimental, Universidade Federal do Rio Grande do Norte, Natal - RN 59072-970, Brazil}
\affiliation{$^{3}$Escola de Ciência e Tecnologia, Universidade Federal do Rio Grande do Norte, 59072-970, Natal- RN, Brazil}

\date{\today}

\begin{abstract}
Significative enhance in magnetocaloric effect due to finite size and surface effects is
reported in Terbium(Tb) thin films in the helimagnetic phase (corresponding to a temperature range from $T_C$=219 K to $T_N$=231 K), for external fields of the order of kOe.
For a Tb thin film of 6 monolayers submitted to an applied field ($\Delta$H= 30 kOe, $\Delta$H= 50 kOe and $\Delta$H= 70 kOe) we report a significative change in adiabatic temperature, $\Delta$T/$\Delta$H, near
the Néel temperature, of the order ten times higher than that observed for Tb bulk.
On the other hand, for small values of the magnetic field, large thickness effects are found. For external
field strength around few kOe, we have found that the thermal caloric efficiency increases remarkably for ultrathin films.
For an ultrathin film with 6 monolayers, we have found $\Delta$T/$\Delta$H=43 K/T while for thicker films, with 20 monolayers, $\Delta$T/$\Delta$H=22 K/T. Our results suggest that thin films of Tb are a promising material for magnetocaloric effect devices for applications at intermediate temperatures.
\end{abstract}



\maketitle


\section{Introduction}
The magnetocaloric effect (MCE) is a phenomenon that enables the temperature of a material to be altered by the application of external fields. Specifically, a magnetic field acting on a substance alters its magnetic state and, consequently, it changes the internal magnetic energy, so that heat is lost or absorbed reversibly. Under adiabatic conditions, when a magnetic field is switched on or off rapidly, the release or absorption of heat is manifested as an increase or reduction in temperature. Recently, the MCE is attracting the interest of both physicists and engineers: physicists because of potential applications in the study of interactions and changes of magnetic structures in magnetic materials, whereas engineers
are hoping to be able to construct new devices and cooling systems\cite{Casey1}. Also, due to the discovery of the giant MCE
near room temperature in bulk Gd$_5$(Si$_x$Ge$_{1-x}$),  by Pecharsky and Gschneidner's\cite{Pecharsky1,Pecharsky2}, the interest in the magnetocaloric effect was sparked.  This made the prospect of commercializing magnetic cooling tenable. Numerous magnetocaloric materials have been studied to
date in bulk form\cite{Phan,Singh,Bruck,Lingwei}, silicides, transition metal
intermetallics, lanthanides, Heusler alloys, and manganites.
In contrast, investigating the magnetocaloric effect and
related materials remains a novel endeavor for nanoscience or in the limit of ultrathin films.
Indeed, exploring the magnetocaloric effect via nanostructuring or ultra thin films,
falls securely within the grand challenges of
nanomagnetism and their frontiers.

In this context, it is the aim of this work to contribute in some way to elucidate the studies of MCE in ultrathin films of rare earth materials. Here we restrict our study to Terbium, due to its large magnetic momentum and interesting magnetic phase
transitions (heat capacity of a magnetic material usually shows a peak at its transition temperature)\cite{Drulis}. In a general way, heavy rare earth elements
and their compounds are considered to be the best-suited materials for achieving a large MCE\cite{Tishin}.
Rare-earth metals have many different magnetic structures resulting from the competition between the crystal field and exchange
interactions. When a magnetic field is applied it gives rise to a third interaction and the magnetic structures are more complicated\cite{mello1,mello2,mello3}. In the
absence of an applied magnetic field, for example, the Terbium orders in a basal plane helical phase in the temperature interval from 231 K (Néel temperature)
to 219 K (Curie temperature). The helix turn angle $\delta\phi$\cite{Satoru} varies between $20.5^\circ$  and $17^\circ$. Therefore the helix period
corresponds to nearly eighteen atomic layers and one might expect strong surface effects for films in this thickness range. When a magnetic field
is applied in the basal plane other magnetic structures are observed including a basal plane ferromagnetic phase, a fan phase and a helifan phase\cite{Jensen}.
In thin films, when the thickness is comparable to the periodicity of the ordered structure, it is expected that even the magnetic arrangement itself can be strongly modified.
Rare earth helimagnets such as Ho, Dy, and Tb represent the best candidates to put into evidence such finite-size effects. This finite-size effect is caused by
the reduced number of atoms in the direction perpendicular to the film surface that leads to a decrease of the total magnetic exchange energy.

The MCE is characterized by a variation $\Delta$T in the temperature of the magnetic material corresponding to an adiabatic change in the external magnetic field intensity.
$\Delta$T  is a function of the initial temperature, T, and of the change in the strength of the external magnetic field, $\Delta$H=H$_{f}$-H$_{i}$, H$_{f}$ and H$_{i}$ are the final and
initial values of the external field, and a positive MCE corresponds to having $\Delta$T$>0$ for $\Delta$H$>0$. Alternatively the MCE may be seen as the change in the magnetic
entropy in an isothermal change of the external field intensity.

For ferromagnetic materials the reduction of the Zeeman energy splitting of the energy levels results from a combination of the direct effect of reducing the external field, and
an indirect effect due to the reduction of the thermal value of the magnetic moment in the field direction. The internal fields, due to exchange energy, are extremely large at
low temperatures, and the external field affects the magnetic order most significantly in the neighborhood of the Curie temperature. In the ordered phase $\Delta$T is an increasing
function of the temperature, with a maximum at the Curie temperature\cite{pecha}. Helimagnetic materials have a rich phase diagram and a complex MCE. In the magnetic ordered temperature
interval the external field may induce changes in the magnetic structure, without appreciable changes in the thermal average value of the magnetic moment per atom. The effect of external
field on the magnetic entropy is dependent upon the field strength, and both positive and negative MCE may occur.

The adiabatic temperature change can be measured directly or indirectly by using heat capacity and magnetization data\cite{pecha2}. In a material displaying MCE, the alignment of randomly oriented
magnetic moments by an external magnetic field results in heating of the material. If the magnetic field is subsequently turned off, the
magnetic moments randomize again, which leads to cooling of the material. The MCE of the Gd$_{5}$(Si,Ge)$_{4}$\cite{Morellon,Moore}
have been extensively studied due to their potential use for magnetic refrigeration applications near room temperature, as we have pointed above.
These compounds undergo a first-order phase transition and exhibit large MCE. Measurements of the MCE in polycrystalline Ho\cite{Nikitin}
displayed MCE maxima at T$_{C}$=20 K ($\Delta$T=4.6 K for $\Delta$H=60.2 kOe) and at T$_{N}$=132 K($\Delta$T=4.5 K for $\Delta$H=60.2 kOe).
Recent works demonstrated that thin Dy\cite{vdm1} and Ho\cite{vdm2} films display MCE of the same order of magnitude found in giant magnetocaloric effect (GMCE) materials.
The enhancement of the MCE, relative to the bulk Dy and Ho values, is more pronounced for small values of the external field strength, where
in ultra-thin films, the helical state forms and the film behaves as a ferromagnetic.

In this paper we are interested to investigate the thickness influence on MCE of Terbium thin films, in the frontier between nanometric (6 to 20 monolayers) and  bulk structures, in the temperature range from 220 K to 230 K. The plan of this paper is: In section \ref{sec1} we present the theoretical model used here. Section\ref{sec2} is devoted to present the results and a discussion about the MCE in Terbium thin films. Finally, is section\ref{sec3} we present the conclusions of this paper.

\section{Theoretical Model} \label{sec1}

We investigate a c-axis thin film, consisting of a stacking of atomic layers with equivalent spins, infinitely extended in the $x$-$y$
directions. The spins in each monolayer are exchange coupled with the spins in the first and second neighbour monolayers. The anisotropy is uniform throughout the film and the near surface spins
have reduced exchange energy. The magnetic Hamiltonian is given by:

\begin{eqnarray}\label{Eq.1}
\mathcal{H}=J_{1}(g-1)^{2}\sum_{n=1}^{N-1}\vec{J}(n)\cdot\vec{J}(n+1)+\nonumber \\
J_{2}(g-1)^{2}\sum_{n=1}^{N-2}\vec{J}(n)\cdot\vec{J}(n+2)+\\
\sum_{n=1}^{N}\left[K_{6}^{6}(T)\cos(6\phi_n)-g\mu_B\vec{J}(n)\cdot\vec{H}
\nonumber\right]
\end{eqnarray}

In Eq. (\ref{Eq.1}),  $J_{1}$ and $J_{2}$ describe the exchange interaction between the nearest and next-nearest monolayers respectively, $\vec{J}$(n) denotes the total angular momentum per atom in the $n$-th monolayer. The coefficient $K_{6}^{6}$(T) describes the hexagonal anisotropy and the last term is the Zeeman Energy, where the external field $\vec{H}$ is applied in one easy direction in the hexagonal plane, making an angle of $30^\circ$ with $x$ axis.

We use the Tb bulk energy parameters\cite{Tishin2}, where the modulus of the total angular momentum is J = 6, J$_1$ = 47$k_B$, J$_2$ = $-J_1$/4$\cos\phi(T)$, where $\phi(T)$ is the temperature dependent helix turn angle \cite{dietri}, g=3/2 is the Landé factor, corresponding to a saturation magnetic moment per atom of 9.7$\mu_B$, and $K_6^6 (T)$  is adjusted so as to reproduce the temperature dependence\cite{Coqblin} of the hexagonal anisotropy energy.

We use a self-consistent local field model which incorporates the surface modifications in the exchange field and the thermal average values $(\langle J(n) \rangle $; $n=1...N)$ and the orientation of the spins in each layer $(\langle \phi_{n} \rangle $; $n = 1...N)$\cite{CamleyRE1,CamleyRE2}. However, the surface effects are not necessarily restricted to the surface layers. The number of layers modified by surface effects depends on the way the effective local field relaxes towards the bulk pattern in the middle of the film.

We calculate the MCE numerically as a function of the magnetization. If the magnetization is a continuous function of the temperature and magnetic field, then the adiabatic temperature change are given by:

\begin{eqnarray}\label{Eq.3}
\Delta {T_{ad}}= - \int_{{H_i}}^{{H_f}}{{{\frac{T}{{C(T,H)}}}_H}{{\left({\frac{{\partial M(T,H)}}{{\partial T}}}\right)}_H}dH}
\end{eqnarray}

\noindent From Eq.\ref{Eq.3} it is easy to state that a material should have large MCE when ${{\left({\frac{{\partial M(T,H)}}{{\partial T}}}\right)}_H}$ is large and C(T,H) is small at the same temperature\cite{Bruck2}.

Following the processes with $H_{i}=0$ and use $\Delta$$T_{ad}(T,H)$ with $H=H_{f}$. Here ${{\left({\frac{{\partial M(T,H)}}{{\partial T}}}\right)}_H}$
is the temperature derivative of the magnetization along the external field direction
and $C(T,H)$ is the heat capacity expressed by\cite{gsch2}:

\begin{eqnarray}\label{Eq.4}
C(T,H)= T \left(\frac{\partial S(T,H)}{\partial T}\right)_H
\end{eqnarray}

\noindent
where the total entropy $S(T,H)$ of a magnetic material is, at constant pressure, the sum of the lattice, electronic and magnetic contributions:

\begin{eqnarray}\label{Eq.5}
S(T,H)= S_{Mag}(T,H)+S_{Latt}(T,H)+S_{Elet}(T,H)
\end{eqnarray}

In this work we ignore the field dependence of the electronic entropy (due to its small contribution to the total entropy). Thus the magnetic contribution is given by:

\begin{eqnarray}\label{Eq.6}
S_{Mag}=\frac{1}{T}\sum_{s=1}^{2J+1}\frac{\mathcal{H}e^{-\mathcal{H}/k_{B}T}}
{Z}+k_{B}T\ln Z
\end{eqnarray}

where $Z=\sum{e^{-\mathcal{H}/k_{B}T}}$ is the partition function, $\mathcal{H}$ is the Hamiltoninan  that describes the energy of the system, given by Eq.\ref{Eq.1}, T is the temperature and $k_{B}$ is the Boltzmann constant. The lattice contribution is given by:

\begin{eqnarray}\label{Eq.7}
S_{Latt}= -3R\ln[1-e^{\Theta_{D}
/T}]+12R\left(\frac{T}{\Theta_{D}}\right)^{3}\int_{0}^{\Theta_{D}
/T}\frac{x^{3}dx}{e^{x}-1}
\end{eqnarray}

\noindent
where {\it R} is the gas constant and $\Theta_{D}$ is the Debye temperature, for Tb the Debye temperature is $\Theta_{D}$=158 K. For a review on the magnetocaloric effect with special attention to nanoscale thin films and heterostructures see ref.\cite{WCasey,Gschneidner}.

\section{Results and Discussion}\label{sec2}

In this section we show the numerical results of the MCE ($\Delta T_{ad}(K)$) for different applied fields and thicknesses of Tb thin films.
Our studies revealed that the efficiency of the MCE in nanofilms of Tb has a strong dependence with its thickness. We note, too, that dependence of field in MCE decreases continuously with increasing thickness of film, approaching the standard value for bulk when the film becomes too thick (in a nanometric scale).

\begin{figure}[!htb]
\centering {\includegraphics[scale=0.3]{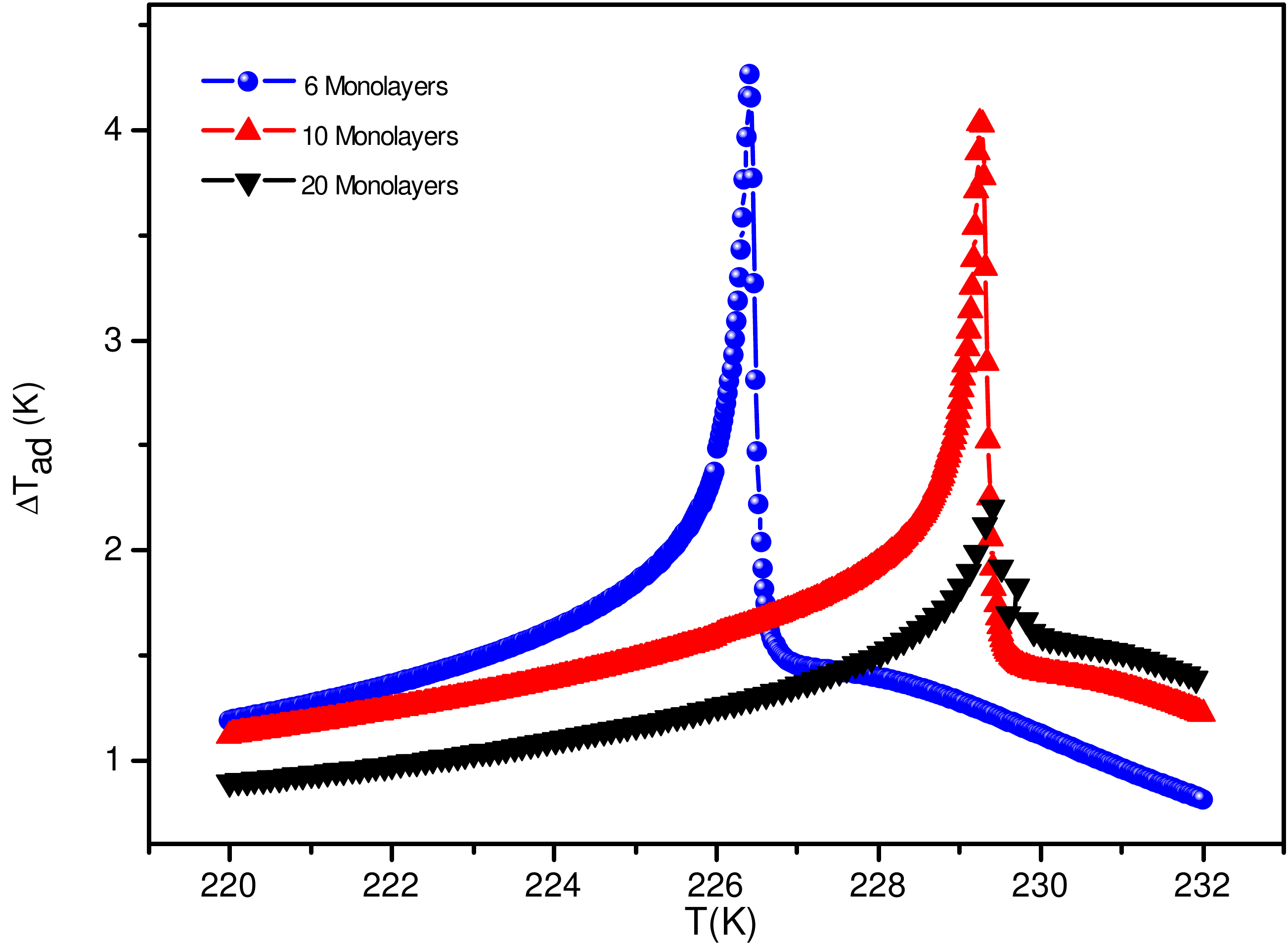}}
\caption{Adiabatic variation of the temperature as a function of T
for a Terbium film with thickness of 6, 10 and 20 monolayers, for an
applied field of $\Delta$H=1 kOe.}\label{fig1}
\end{figure}

We show on Fig. \ref{fig1} the variation of $\Delta T_{ad}(K)$ for films of Tb with 6, 10 and 20 monolayers for an external field of $\Delta H$=1 kOe.
We have chosen a small value of H because the thin film effects on the MCE properties are more visible for small values of the external field.
We found significant $\Delta T(T,H)$/$\Delta H$ values range from 43 $K/T$ to 6 monolayers thin film, 40 $K/T$ to 10 monolayers thin film and 22 $K/T$ to 20 monolayers thin film.

\begin{figure}[!htb]
\centering {\includegraphics[scale=0.3]{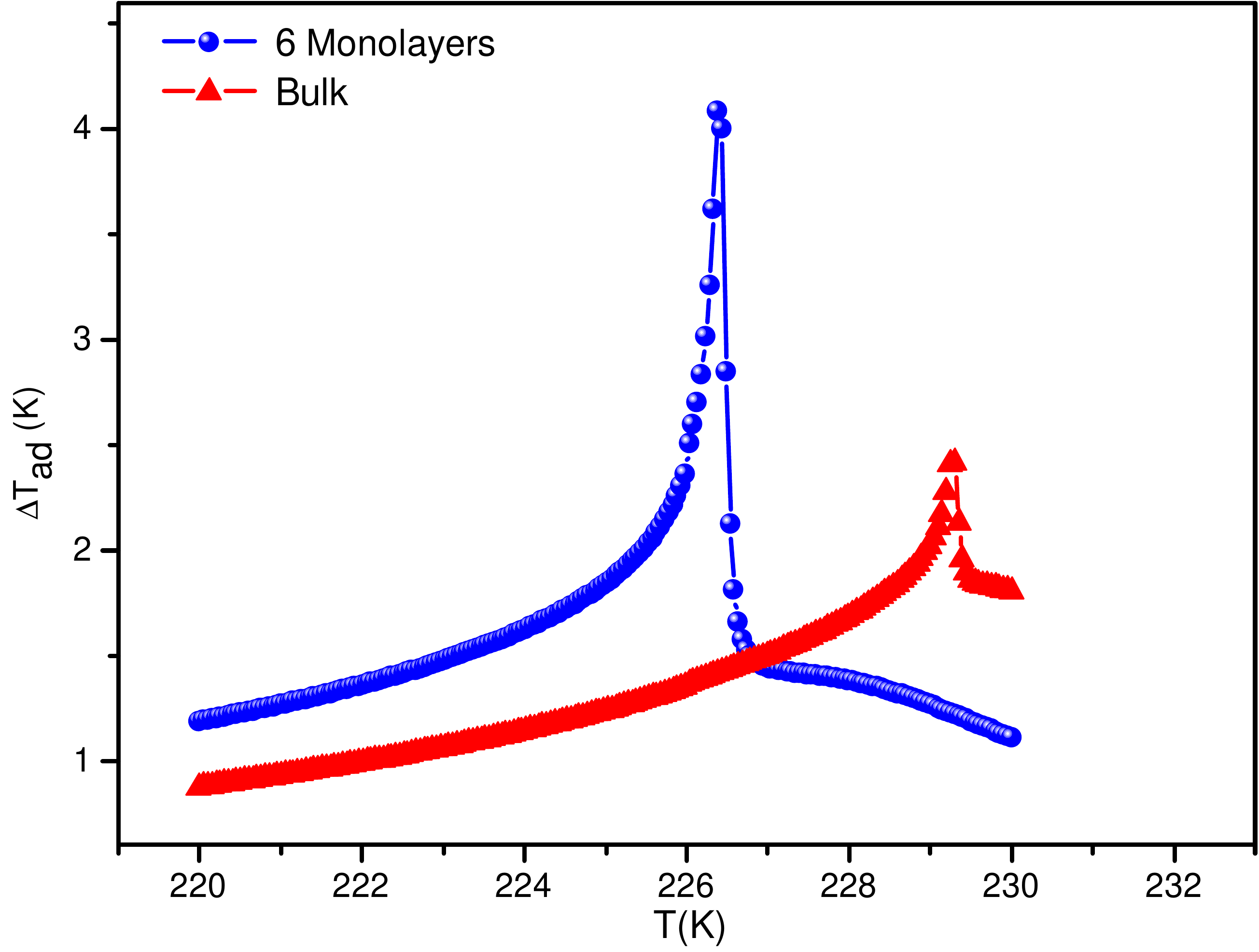}}
\caption{Adiabatic variation of the temperature as a function of T
for a Terbium film with thickness of 6 monolayers and bulk, for an
applied field of $\Delta$H=1 kOe.}\label{fig2}
\end{figure}

In order to stress the thickness influence on the MCE, we show on Fig. \ref{fig2} the variation of $\Delta T_{ad}(K)$ for a 6 monolayer Tb film and for bulk Tb, for an external field $\Delta H$=1 kOe. We found that the peak value of $\Delta T_{ad}(K)$ for the thin film is nearly two times larger than the corresponding bulk value.

We can also observe, Fig. \ref{fig1} and Fig. \ref{fig2}, the dependence of the Néel temperature with the film thickness, where for thick films it approximates Bulk $T_N$ and for thinner films the temperature decreases.
Near the Néel temperature the bulk and thin film values of the helix turn angle are similar, $\phi(T)\approx 20^\circ$. Thus a complete helix corresponds to around eighteen atomic layers. One might then expect to find bulk MCE properties for films with more than 20 monolayers.

The key feature for thin film effects is the field dependence of the field derivative of the adiabatic temperature rise, as seen from Eq.\ref{Eq.3}, $(dT/dH)_{H}$=$-$T($\partial M(T,H)/\partial T)_H$/C(T,H)$_{H}$. The number of modified layers by the surface effect depends on the way the local effective field relaxes the spins in the middle of film. In this manner, changes will occur in magnetic phases of thin films, compared with the bulk and consequently on the values of ($\partial M(T,H)/\partial T)_H$, significantly affecting the value of $\Delta T_{ad}$. As a result, the adiabatic temperature change is larger than the corresponding value for bulk.

\begin{figure}[!htb]
\centering {\includegraphics[scale=0.3]{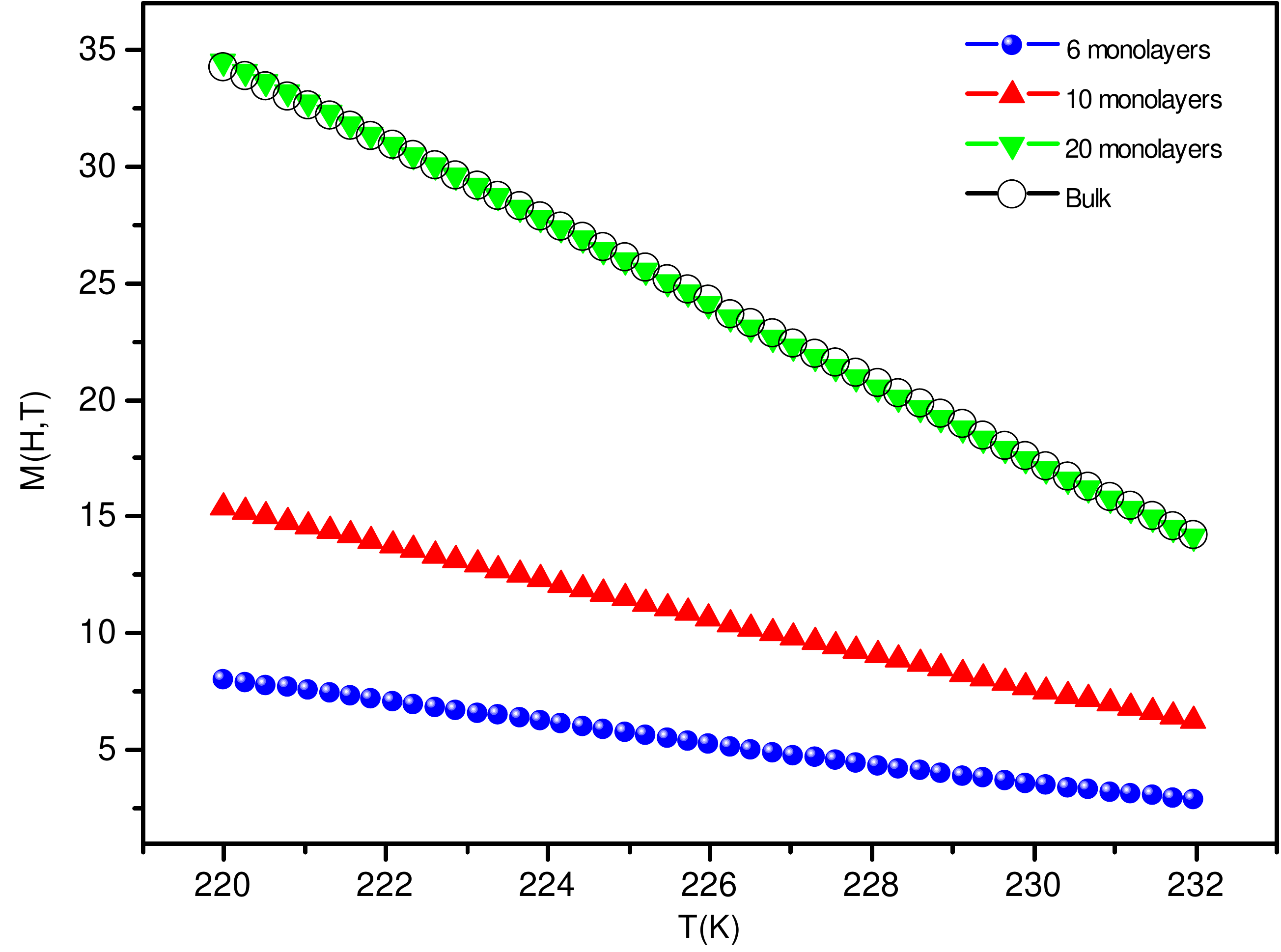}} \caption{Isofield
magnetization curves for external field H=1 kOe. The magnetization
is shown in units of $µ$B.}\label{fig3}
\end{figure}

\begin{figure}[!htb]
\centering {\includegraphics[scale=0.3]{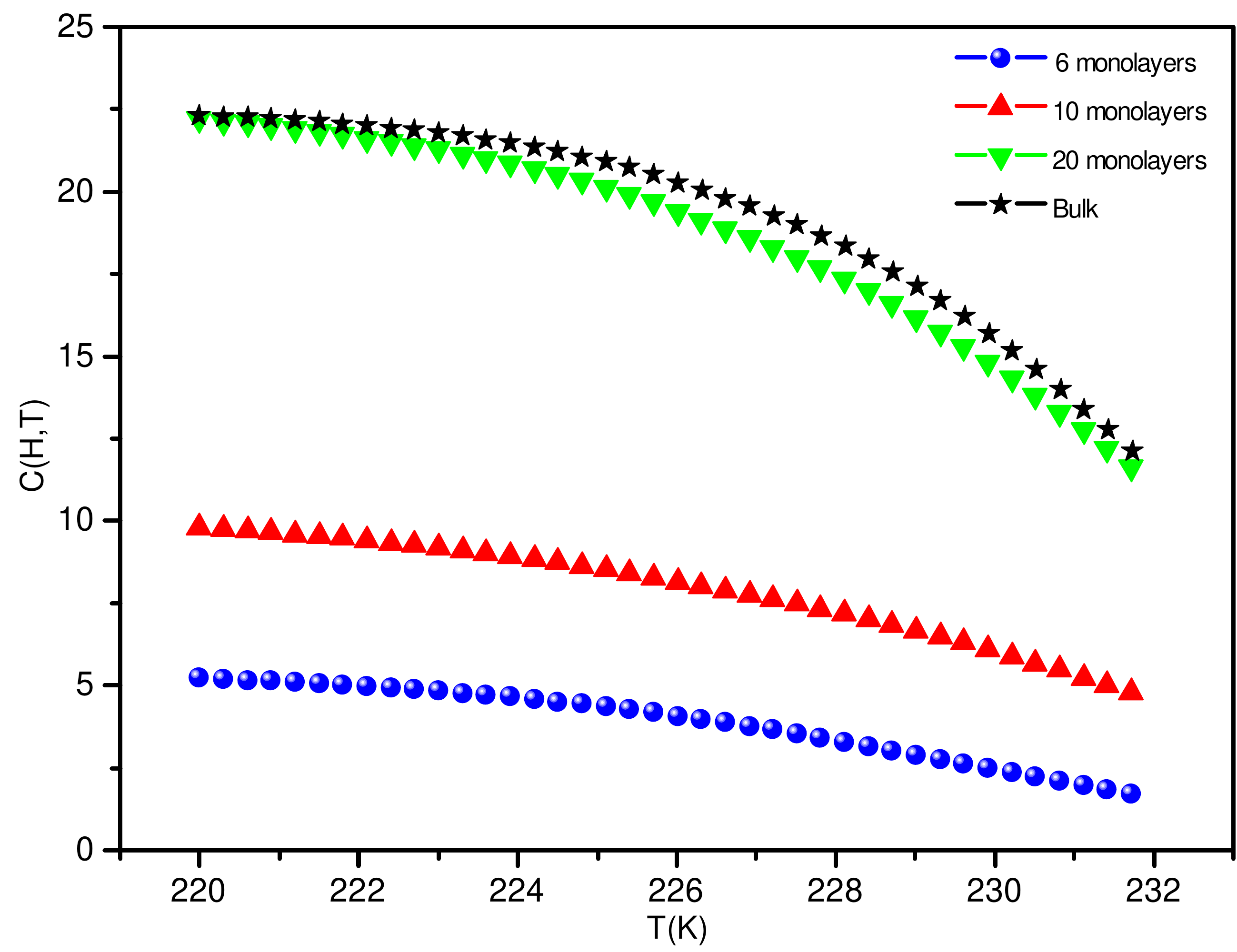}} \caption{C(H,T)
for an applied field of $\Delta$H=1 kOe.}\label{fig4}
\end{figure}

On Fig. \ref{fig3} and Fig. \ref{fig4} we show the contribution of isofield magnetization and heat capacity for $\Delta T_{ad}(K)$ for films of Tb with 6, 10, 20 monolayers and bulk, for an external field of $\Delta H$=1 kOe, where we can observe that for thicker films, these contributions resemble those of bulk. This fact makes the $\Delta T_{ad}(K)$ of thick films close to that of volume, imposing a theoretical and practical approximation for MCE simulations in Tb nanofilms.

\begin{figure}[!htb]
\centering {\includegraphics[scale=0.3]{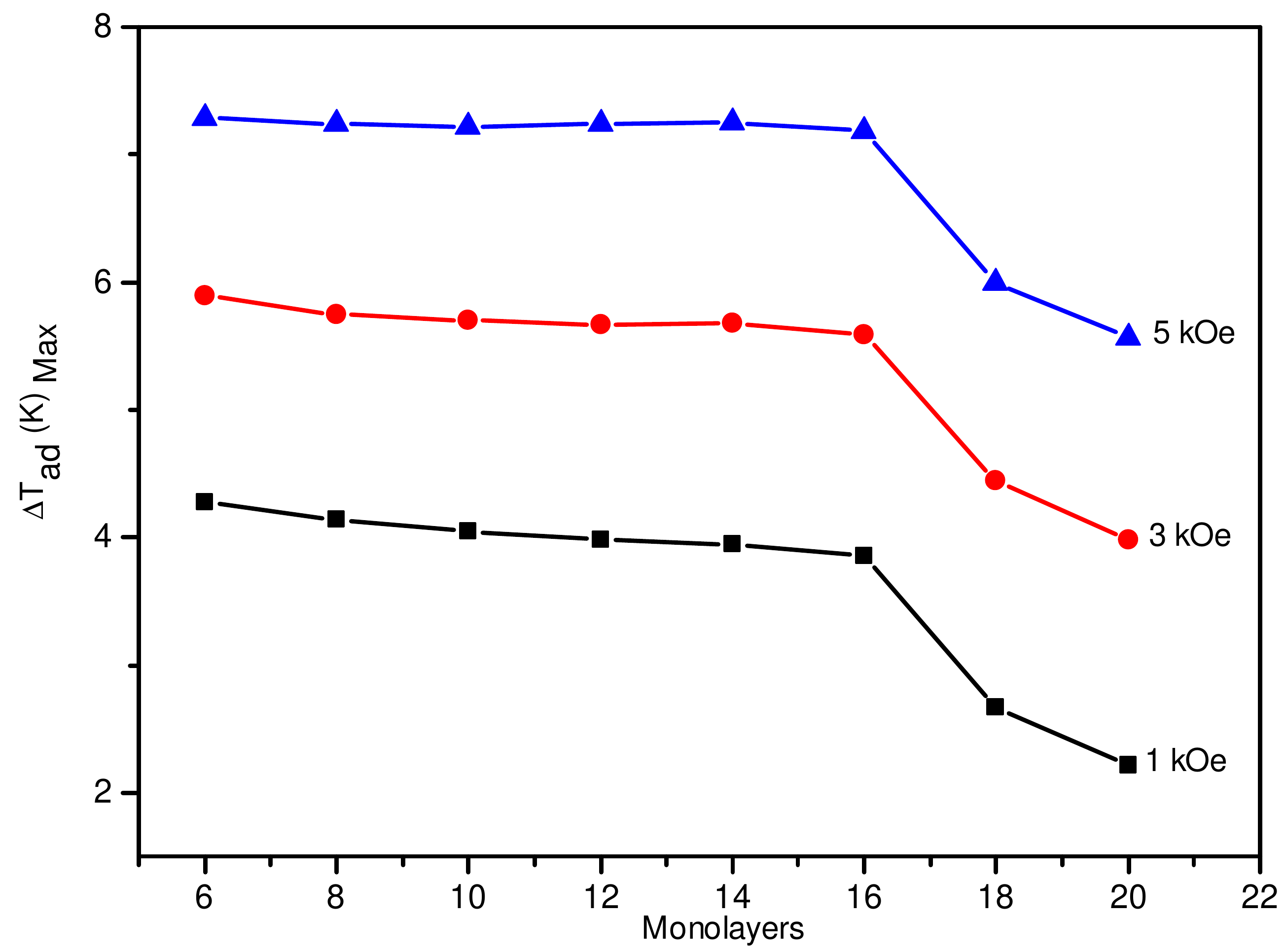}}
\caption{Monolayers dependence of the value $\Delta T(T,H)_{Max}$
for an external field $\Delta H$=1 kOe, $\Delta H$=3 kOe and $\Delta
H$=5 kOe}\label{fig5}
\end{figure}

We show in Fig. \ref{fig5} that the MCE efficiency, $\Delta T(T,H)$/$\Delta H$, increases as the thickness is reduced. For instance, for a thin film of 6 monolayers, we have approximately: 43 $K/T$, 19 $K/T$ and 14 $K/T$ for external applied fields of $\Delta H$=1 kOe, $\Delta H$=3 kOe and $\Delta H$=5 kOe, respectively. For a thin film of 18 monolayers, we will have efficiencies of approximately: 26 $K/T$, 15 $K/T$ and 12 $K/T$
for external applied fields of $\Delta H$=1 kOe, $\Delta H$=3 kOe and $\Delta H$=5 kOe, respectively. Therefore, the finite size enhancement of the MCE is larger for a small field strength.

\begin{figure}[!htb]
\centering {\includegraphics[scale=0.3]{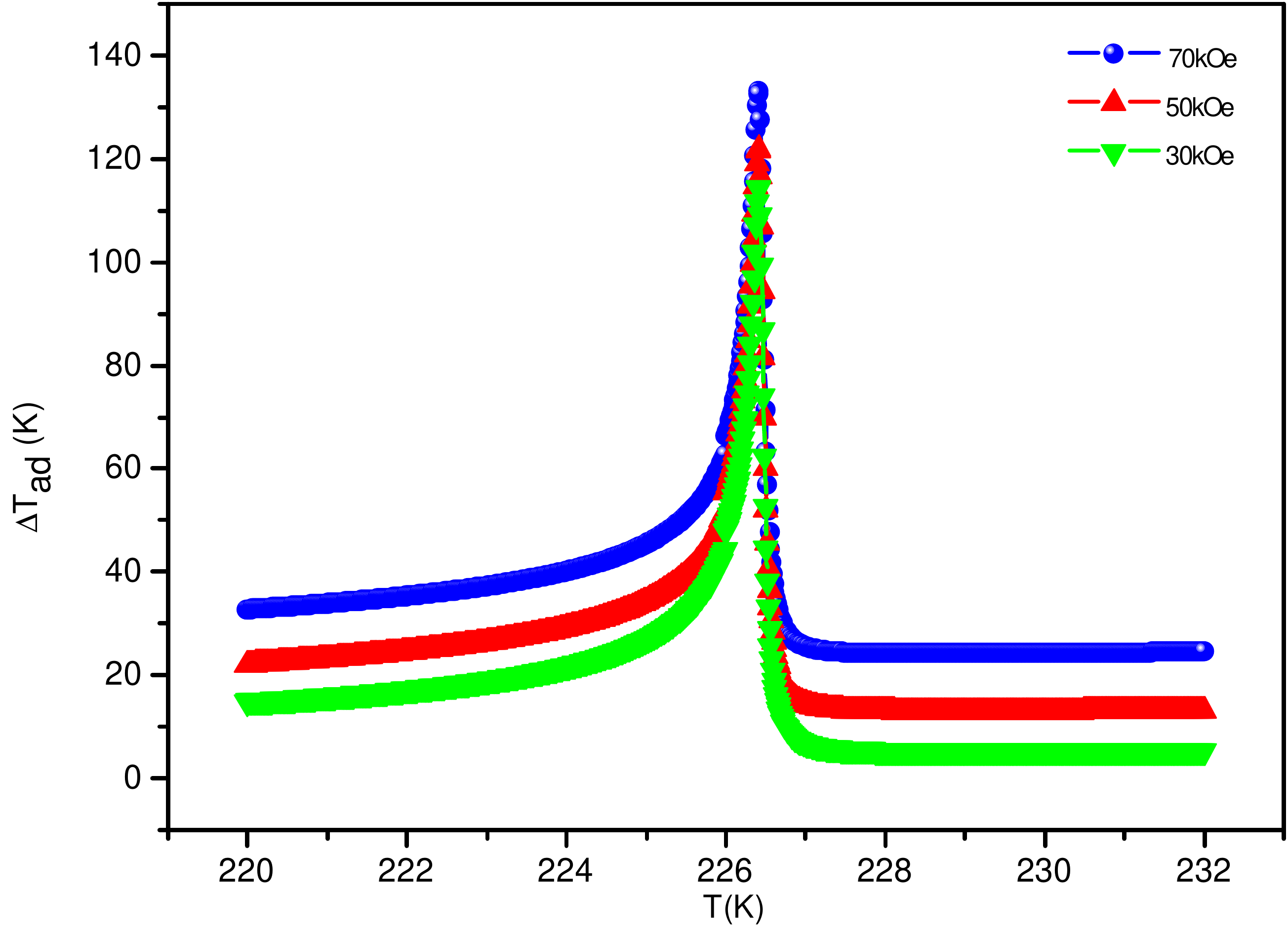}}
\caption{Adiabatic variation of the temperature as a function of T
for a Terbium film with 6 monolayers for applied fields
($\Delta$H=30 kOe, $\Delta$H=50 kOe and $\Delta$H=70
kOe).}\label{fig6}
\end{figure}

On Fig. \ref{fig6} we depict the adiabatic temperature variation for a Tb film of 6 monolayers subjected to external fields (30 kOe, 50 kOe and 70 kOe), with values of $\Delta T(T,H)$/$\Delta H$ (39 $K/T$, 25 $K/T$ and 19 $K/T$) respectively, where we observed that the dependence of the MCE on the applied field is much stronger than the obtained for bulk Tb. Near the Néel temperature, the value found for $\Delta T_{ad}$ is approximately ten times higher than that obtained for polycrystalline samples of Tb\cite{Tishin2}. This effect is referred as giant magnetocaloric effect (GMCE), a name proposed by Pecharsky and Gschneidner in 1997 as they investigate Gd$_{5}$Si$_{2}$Ge$_{2}$ \cite{pecha2}. From the previous results we can conclude that Tb thin films are indeed a promising magnetocaloric material for applications at intermediate temperatures.

\section{Conclusions} \label{sec3}

In conclusion, the present study demonstrated that the MCE efficiency of Tb is significantly enhanced by
finite size and surface effects. Confinement in rather thin films, with thickness below the helix period, favors a ferromagnetic state.
This is due to the near surface spins are more easily turned in the direction of the external field. Near the surfaces the turn angle is smaller,
due to the lack of second neighbors which favors a  ferromagnetic-like configuration in the whole temperature range from the Curie temperature to
the Néel temperature, leading to a giant MCE. The present results suggest that thin films of Tb might be a promising material of MCE
devices for applications at intermediate temperatures, for example, the cooling natural gas to liquid form, so that it can be cost-effectively shipped long distances.
Finally, we hope that our results can stimulate the experimentalists to probe the findings presented here.

\begin{acknowledgments}
The authors acknowledge the financial support from the Brazilian research agencies CNPq and FAPERN.
\end{acknowledgments}


\begin{references}

\bibitem{Casey1} Casey W. Miller, D.V. Williams, N.S. Bingham, H. Srikanth, J. App. Phys. {\bf 107}, 09A903 (2010).

\bibitem{Pecharsky1} V.K. Pecharsky and K.A. Gschneidner, Jr., Phys. Rev. Lett. {\bf 78}, 4494 (1997).

\bibitem{Pecharsky2} V.K. Pecharsky and K.A. Gschneidner, Jr., Appl. Phys. Lett. {\bf 70}, 3299 (1997).

\bibitem{Phan} M.H. Phan and S.C. Yu, J. Magn. Magn. Mater. {\bf 308}, 325 (2007).

\bibitem{Singh} N.K. Singh, K. Suresh, A. Nigam, S. Malik, A. Coelho, and S. Gama, J. Magn. Magn. Mater. {\bf 317}, 68 (2007).

\bibitem{Bruck} E. Br\"uck, O. Tegus, D.C. Thanh, N.T. Trung, and K. Buschow, Int. J. Refrig. {\bf 31}, 763 (2008).

\bibitem{Lingwei} Li. Lingwei, Yi. Yalin, Su. Kunpeng, Qi. Yang, Huo. Dexuan and R. P\"ottgen, J. Mater. Sci. {\bf 51}, 5421 (2016).

\bibitem{Drulis} H. Drulis, A. Hackemer, L. Folcik, A. Zaleski, Solid State Commun. {\bf 150}, 164 (2010).

\bibitem{Tishin} A.M. Tishin, J. Appl. Phys. {\bf 68}, 6480 (1990).

\bibitem{mello1} V.D. Mello and A.S. Carriço, Surf. Sci. {\bf 482}, 960 (2001).

\bibitem{mello2} F.H.S. Sales, A.L. Dantas, V.D. Mello and A.S. Carriço, J. Mater. Sci. {\bf 45}, 5036 (2010).

\bibitem{mello3} L.J. Rodrigues, V.D. Mello, D.H.A.L. Anselmo, M.S. Vasconcelos, J. Magn. Magn. Mater. {\bf 377}, 24 (2015).

\bibitem{Satoru} S. Kobayashi, Phys. Rev. B {\bf 84}, 224418 (2011).

\bibitem{Jensen} J. Jensen and A.R. Mackintosh, in \textit{ Rare Earth  Magnetism}, (Oxford University Press, Oxford, 1991).

\bibitem{pecha}  V.K. Pecharsky and K.A. Gschneidner, J. Magn. Magn.  Mater. {\bf 200}, 44 (1999).

\bibitem{pecha2} V.K. Pecharsky and K.A. Gschneidner, J. Appl. Phys. {\bf 86}, 565 (1999).

\bibitem{Morellon} L. Morellon, J. Blasco, P.A. Algarabel and M. R. Ibarra, Phys. Rev. B {\bf 62}, 1022  (2000).

\bibitem{Moore}  J.D. Moore, G.K. Perkins, Y. Bugoslavsky, L.F. Cohen, M.K. Chattopadhyay, S.B. Roy, P. Chaddah, K.A. Gschneidner, Jr., and V.K. Pecharsky, Phys. Rev. B {\bf 73}, 144426 (2006).

\bibitem{Nikitin} S.A. Nikitin, A.S. Andreenko, A.M. Tishin, A.M. Arkharov and A.A. Zherdev, Phys. Met. Metallogr. {\bf 60}, 56 (1985a).

\bibitem{vdm1} V.D. Mello, A.L. Dantas and A.S. Carri\c{c}o, Solid State Commun. {\bf 140}, 447 (2006).

\bibitem{vdm2} F.C. Medeiros, V.D. Mello, A.L. Dantas, F.H.S. Sales and A.S. Carri\c{c}o, J. Appl.  Phys. {\bf 109}, 07A914 (2011).

\bibitem{Tishin2} A.M. Tishin and Y.I. Spichkin, in \textit{The Magnetocaloric  Effect and its Applications}, (IOP Publ. Ltd., Bristol, 2003).

\bibitem{dietri} O.W. Dietrich and J. Als-Nielsen, Phys. Rev. {\bf 162}, 315 (1967).

\bibitem{Coqblin} B. Coqblin, in \textit{The Electronic Structure of Rare-Earth Metals and Alloys: the Magnetic Heavy Rare-Earths}, (Academic Press, New York, 1977).

\bibitem{CamleyRE1} R.E. Camley, Phys. Rev. B {\bf 35}, 3608 (1987).

\bibitem{CamleyRE2} A.L. Dantas, R.E. Camley and A.S. Carriço, Phys. Rev. B {\bf 75}, 094436 (2007).

\bibitem{Bruck2} E. Bruck, J. Phys.D: Appl. Phys {\bf 38}, R381 (2005).

\bibitem{gsch2} K.A. Gschneidner Jr, V.K. Pecharsky and A.O. Tsokol,  Rep. Prog. Phys. {\bf 68}, 1479 (2005).

\bibitem{WCasey} W. Casey, D.D. Belyea, B.J. Kirby, J. Vacu. Sci. \& Tech. A {\bf 32}, 040802 (2014).

\bibitem{Gschneidner} K.A. Gschneidner, Jr. and V.K. Pecharsky, Annu. Rev. Mater. Sci. {\bf 30}, 387 (2000).

\end{references}
\end{document}